\documentstyle[preprint,aps]{revtex}
\input psfig
\newcommand{\psplot}[1]{ \hbox to\textwidth{\hfill
             \psfig{figure=#1.ps,width=\textwidth}\hfill} }

\def\rnoch{r_{c\!\!/}}%

\begin{document}

\def\ltap{\;\raisebox{-.4ex}{\rlap{$\sim$}} \raisebox{.4ex}{$<$}\;}
\def\gtap{\;\raisebox{-.4ex}{\rlap{$\sim$}} \raisebox{.4ex}{$>$}\;}

{\tighten
\preprint{\vbox{\hbox{FERMILAB--PUB--95/167--T}
                \hbox{HUTP-95/A009}
                \hbox{\today}}}
\title{Hadronization of $b\rightarrow c\bar cs$}

\author{Gerhard Buchalla, Isard Dunietz}
\address{\it Fermi National Accelerator Laboratory, P.O.~Box 500,
Batavia, Illinois 60510}
\author{and}
\author{Hitoshi Yamamoto}
\address{\it Harvard University, Cambridge, MA 02138}


\maketitle
\begin{abstract}
The $b\rightarrow c\bar cs$ transition is usually believed to
hadronize predominantly
in $\overline B \rightarrow X_c D^{(*)-}_s$ with the $D^{(*)-}_s$
originating
from the virtual $W$.  We demonstrate in a variety of independent
ways that other
hadronization processes cannot be neglected. The invariant mass of
$\bar c s$ has sizable phase-space beyond $m_D +m_K$. The rate for
$\overline B
\rightarrow D\overline D \;\overline K X$ could be significant and
should not be ignored
as was done in previous experimental analyses.  We estimate the
number of charmed hadrons per $B$-decay, $n_c$, to be $\approx 1.3$
to higher accuracy than obtained in previous investigations.  Even
though $n_c$ is currently measured to be about $1.1$, observing a
significant $\overline B
\rightarrow D\overline D \;\overline K X$ would support $n_c \approx
1.3$.  Many testable consequences result, some of which we discuss.

\end{abstract}

\pacs{}
}

At present, there appears to be a conflict between experiment and
theory for
fitting both the inclusive semileptonic branching ratio and the
number
of charmed hadrons per $B$ decay
\cite{altarelli,baffle,fwd,bagan1,bagan2,bagan3,voloshin,pub361},
\begin{equation}
\label{nc}
n_c = 1 - B(b\rightarrow \mbox{no charm} )+ B(b\rightarrow c\bar c
s^\prime ) \approx 1 + B(b\rightarrow c\bar c s^\prime )\;.
\end{equation}
The prime indicates that the corresponding Cabibbo-suppressed mode is
included.  Experimentally the inclusive semileptonic $BR$ has been
measured accurately to be
\cite{browder}
\begin{equation}
\label{brsl}
B(\overline B\rightarrow X\ell\nu ) =(10.4 \pm 0.4)\% \;,
\end{equation}
and $n_c$ is measured as \cite{browder}
\begin{equation}
\label{ncexp}
n_c = 1.10 \pm 0.06 \;.
\end{equation}

A value of $B(b\rightarrow c\bar c s^\prime ) \approx 0.1$, suggested
by Eqs.~(\ref{nc}) and (\ref{ncexp}), would lead to a theoretical
prediction of $B(\overline B\rightarrow X\ell\nu )$ that is too
large---i.e., inconsistent with its measured value (\ref{brsl}).  On
the other hand, theory predicts $n_c \approx 1.3$ when the observed
semileptonic BR is used as input, which is demonstrated below.  Thus
a conflict arises between (\ref{brsl}) and
(\ref{ncexp})~\cite{baffle,fwd}.

Recently,
Bagan et al.~and Voloshin made progress on
the theoretical side~\cite{bagan1,bagan2,bagan3,voloshin}.
Bagan et al.~\cite{bagan1,bagan2,bagan3}
performed a complete next-to-leading order analysis of inclusive $B$
decays, which included important final state mass effects in the QCD
corrections.  The predicted $B(\overline B\rightarrow X\ell\nu )$
agrees with ~(\ref{brsl}), within uncertainties that are dominated by
renormalization scale-dependences in the perturbative
calculation~\cite{bagan1,bagan2,bagan3}.  Simultaneously, an
enhancement of $B(b\rightarrow c\bar c s^\prime )$ was
found~\cite{bagan1,bagan2,bagan3,voloshin}, albeit with considerable
uncertainties.
Table I summarizes these recent theoretical
findings~\cite{bagan3}.  The main sources of the large errors
in those studies are  dependence on the
renormalization-scale ($m_b / 2 < \mu < 2\; m_b$), dependence
on the renormalization-scheme ($\overline{MS}$ versus pole mass), and
uncertainties in
quark masses.
Although this theoretical analysis hints that $n_c$ may be larger
than
currently
measured \cite{bagan2,bagan3,voloshin,baffle,fwd}, it is difficult to
draw firm conclusions from this direct calculation of $B(b\rightarrow
c\bar c s^\prime )$ in view of the large uncertainties.

It should also
be stressed at this point, that the experimental determination of
$B(\overline B\rightarrow X\ell\nu )$ is reliable and accurate.  In
contrast, the measurement of $n_c$ is a sum over the inclusive yields
of many
charmed hadron species in $B$ decays. It is thus prone to large
uncertainties, perhaps larger than currently realized.

Figure 1 displays the discrepancy graphically.  We discuss now in
some detail how the theoretical curve has been generated. Our
objective is to draw the
most accurate curve of $n_c$ versus semi-electronic $BR$ with
presently
available theoretical calculations.
We do not use the prediction for $B(b\rightarrow c\bar cs')$ because
it
involves large errors, but rather proceed as follows.  We start with
\begin{equation}
\label{bc}
B(b\rightarrow c) = 1 - B(b\rightarrow \mbox{no charm})\;,
\end{equation}
where $B(b\rightarrow \mbox{no charm})$ is small, typically at the
percent level. We take
\begin{equation}
\label{brare}
\rnoch \equiv
\Gamma( b\rightarrow \mbox{no charm})/ \Gamma( b\rightarrow c e\nu) =
0.25 \pm
0.10,
\end{equation}
to account for the small fraction of $b \to s + \mbox{no charm}$
\cite{simma} and charmless $b \to u$ transitions.
Furthermore we use
\begin{equation}\label{ctau}
r_\tau \equiv
\frac{\Gamma(b\rightarrow c\tau\nu )}{\Gamma(b\rightarrow ce\nu )} =
0.25 \; ,
\end{equation}
which is in accordance with the result of Ref.~\cite{falklnn} and
also agrees with a recent ALEPH measurement \cite{aleph},
\begin{equation}
B(b\rightarrow X\tau\nu )= (2.75 \pm 0.30 \pm 0.37 ) \% \; .
\end{equation}
The last required ratio is $\Gamma (b\rightarrow c\bar ud')/\Gamma
(b\rightarrow ce\nu )$ where the dominant uncertainties in
$|V_{cb}|^2$ and in fermion masses cancel.
Bagan et al.~\cite{bagan1} have presented a complete computation of
this quantity in next-to-leading logarithmic approximation taking all
final-state charm quark mass effects into account.  Based on this
perturbative calculation and also including nonperturbative
corrections up to ${\cal O}(1/m_b^2)$, the analysis of~\cite{bagan1}
yields,
\begin{equation}
\label{cud}
r_{ud} \equiv
\frac{\Gamma(b\rightarrow c\bar ud')}
{\Gamma (b\rightarrow ce\nu )}=4.0 \pm 0.4 \;.
\end{equation}
Here the error comes almost entirely from the renormalization-scale
uncertainty and represents a conservative estimate when working to
order ${\cal O}(1/m_b^2)$.
Because nonperturbative effects at ${\cal O}(1/m_b^3)$ could
introduce rate-differences at the $10 \%$ level between $B^-$ and
$\overline B_d$ decays governed by $b\rightarrow c\bar
ud$~\cite{bigistone}, there is considerable room for additional
studies.

Combining Eqs.~(\ref{bc}), (\ref{brare}), (\ref{ctau}),
(\ref{cud}), the $b\to c\bar cs'$ branching fraction can be written
as
\begin{eqnarray}
B(b\rightarrow c\bar cs') &=&
 1 - (2 + r_\tau + r_{ud} + \rnoch)\;B(\overline B\to X_c e\nu )
           \nonumber \\
 &=& 1 - (6.50\pm 0.40) \;B(\overline B\rightarrow X_c e\nu ) \; .
\label{ccsbrse}
\end{eqnarray}
In this relation the very small contribution from $b\rightarrow
u\bar c s^\prime$ transitions has been neglected.
Eqs.~(\ref{nc}) and (\ref{ccsbrse}) yield the number of charms per
$B$ decay as
\begin{eqnarray}
n_c &=& 2 - (2 + r_\tau + r_{ud} + 2 \rnoch)\;
B(\overline B\to X_c e\nu ) \nonumber \\
&=& 2 - (6.75\pm 0.40) \;B(\overline B \rightarrow X_c e\nu ) \;,
\label{ncbrse}
\end{eqnarray}
where we note that $B(b\to c\bar cs')$ drops out in the
linear relation between $n_c$ and $\overline B\to X_c e\nu$, and that
the relation is largely free from uncertainties in masses of $b$ and
$c$
quarks since
the error is dominated by the uncertainty in $r_{ud}$.
Figure 1 shows the discrepancy between theory given by
Eq.~(\ref{ncbrse}) and experiment.

The precisely measured semileptonic $BR$ together with
Eqs.~(\ref{ccsbrse})-(\ref{ncbrse}) gives
\begin{equation}
\label{ccsexpect}
B(b\rightarrow c\bar cs') = 0.32 \pm 0.05 \; ,
\end{equation}
\begin{equation}\label{ncexpect}
n_c = 1.30 \pm 0.05.
\end{equation}
This is our central result.  Our predictions for $B(b\rightarrow
c\bar cs')$ and for $n_c$ agree with the central values obtained in
previous theoretical
investigations~\cite{bagan2,bagan3,voloshin,baffle,fwd}
but have smaller errors.
As discussed in more detail below, such a large value of
$B(b\rightarrow
c\bar cs')$ requires a significant rate for
$\overline B\rightarrow D\overline D \;\overline K X$.
We predict the observation of (a) $\overline B\rightarrow D^{(*)}
\overline
D^{(*)} \overline K$ modes with significant $BR$'s, (b) enhanced
$\ell^+ \overline D$
and $\ell^- D$ correlations where the primary lepton originates from
one
$B$ and the charmed hadron from the other $B$ in the event, and
(c) enhanced $DD$ and $\overline D \;\overline D$ correlations
at the $\Upsilon (4S)\rightarrow B\overline B$.

If the predicted effects will be observed, then the $B(b\rightarrow
c\bar
cs')$
is larger than currently determined by experiment. The measured
number of charm
per $B$ will not change by those observations, but the larger
$B(b\rightarrow c\bar cs'$) would indicate that the current
experimental value of
 $n_c$ is underestimated.  In that case, a careful re-evaluation of
all errors involved
in measuring $n_c$ would be in order, including
re-assessments of absolute $BR$'s of the charmed hadrons some of
which are
poorly known.
On the other hand, non-observation of our predictions would indicate
an enhancement of the $b\rightarrow c\bar ud$ transition over the
parton estimate~\cite{honscheid} and/or a larger rate than
anticipated for charmless $b \to s$
transitions~\cite{palmerstech,kagan}.

Theory alone or experimental measurements alone have large inherent
uncertainties
for $B(b\rightarrow c\bar cs')$.
We therefore adopted a hybrid approach which uses well
measured quantities from
experiment in conjunction with reliably calculated quantities from
theory to
determine
$B(b\rightarrow c\bar cs')$ to higher accuracy \cite{pub361}.

One conventional way to determine $B(b\rightarrow c\bar cs)$ is
to add the inclusive yield of $D_s$ \cite{browder,muheim,palmerstech}
\begin{equation}\label{rds}
R_{D_s} \equiv B(\overline B\rightarrow D^-_s X)+B(\overline
B\rightarrow
D^+_s X)
\end{equation}
to the other observed final states governed by $b\rightarrow c\bar
cs$~\cite{br},
\begin{eqnarray}
B(b\rightarrow c\bar c s)& = & R_{D_s} +B(\overline B\rightarrow
\Xi_c
\overline{\Lambda}_cX)+B(\overline B\rightarrow (c\bar c)X) =
\nonumber \\
& = & 0.12 + 0.01 + 0.03 = 0.16 \pm 0.02 \;.
\end{eqnarray}
$(c\bar c)$ denotes charmonia not seen in $D\overline DX$ such as
$J/\psi , \psi ', \eta_c ,\eta_c^\prime , \chi_c , h_c ,\;
^{1,3}\!\!D_2$.
Within errors, this agrees with the experimental measurement of $n_c$
\begin{equation}
B(b\rightarrow c\bar cs') =n_c - 1 + B(b\rightarrow \mbox{no charm} )
= 0.13 \pm 0.06 \;.
\end{equation}
The agreement appears to support the low value of $n_c$.

Our determination of $B(b\rightarrow c\bar cs')$ suggests a different
picture as to how $b\rightarrow c\bar cs$
hadronizes. A systematic classification shows that five classes of
hadronization can occur, see Table II.
Conventional wisdom \cite{browder,palmerstech,muheim} assumes that
most of
the inclusive $D_s$ production in $B$ decays originates from the
virtual
``W". Motivated by the observed inclusive momentum spectrum of $D_s$
in $B$ decays and by a simple theoretical argument given below,
we predict instead that only about 70\% of the inclusive $D_s$ yield
in
$B$ decays contribute to $\overline B\rightarrow DD^-_s X$ processes.
The remaining $D_s$ (about  30\%) could occur in conjunction with $s\bar s$
fragmentation.
We will return to this point below.

The branching ratio for class (a) is thus
depleted and becomes about $0.7 R_{D_s}$. [This branching ratio
can be at most  $R_{D_s}$, which would soften our conclusion by a
small amount only].
The branching ratios of the observed classes (a)-(c), do not add up
to 30\%.
Thus class (d) must have a sizable branching fraction of about
20\%,
\begin{equation}
B(\overline B\rightarrow D\overline D\;\overline KX) \sim 20\% \;.
\end{equation}
There are several interesting experimental implications. Those modes
can be studied
at CLEO and at LEP.  CLEO has higher statistics, whereas LEP has the
ability to separate one $B$ from the other $b$ hadron. Thus far,
however, they have not been seriously searched for.  The low
$Q$ value in this process suggests that a significant portion will be
three body \cite{nokstar},
$$\overline B\rightarrow D^{(*)} \overline D^{(*)} \overline K \;.$$
Because the responsible Hamiltonian is isospin zero, many isospin
relations can
be used to facilitate the observation of those modes
\cite{dyisospin}.

Finally, the class (e) processes involve $s\bar s$ fragmentation.
Their branching ratio could be non-negligible, at the few percent
level \cite{bdy2}.
A few exclusive final states would then carry the lion's share of the class
(e)
branching ratio, because of limited phase-space.

Before proposing a number of tests, we discuss briefly
a few additional indications that support our hypothesis from

(a) a naive Dalitz plot analysis~\cite{bjorken},

(b) measured inclusive kaon yields in $B$ decays, and

(c) measured inclusive $D$ momentum spectra in $B$ decays.

Figure 2 shows the $b\rightarrow c\bar c s$ Dalitz plot resulting
from
the $(V-A) \times (V-A)$ matrix element, where the initial and final
spins were averaged and summed.
In this simple model, the $\bar cs$
system hadronizes as a $D^-_s X$ for invariant $\bar cs$ masses below
$m_D +m_K$.  In contrast, for
$$m_{\bar cs} > m_D +m_K$$
the $\bar cs$ is not seen as a $D^-_s X$ but rather
as $\overline D \;\overline K X$.
  The Dalitz plot region contributing to
$D_s$ production in $b\rightarrow c\bar cs$ decay is $m_{\bar cs} <
m_D
+m_K$, and one obtains
$$\frac{\Gamma(b\rightarrow c +D^{-}_s )}{\Gamma (b\rightarrow c\bar
cs)}
\approx 0.35 \;.$$

This argument suggests that a large fraction of $b\rightarrow c\bar
cs$
transitions has not been accounted for in previous
investigations~\cite{browder,muheim}. (See however the
analyses of Refs.~\cite{bjorken,ter} which reach
similar conclusions to ours.)  Of course, the naive Dalitz plot
argument is rather crude.  It does not address
issues of hadronization, resonance bands and their interferences,
QCD-corrections, and interferences between penguin-amplitudes
$(b\rightarrow s)$ with the dominant spectator-amplitude
$(b\rightarrow c\bar cs$).
Nevertheless, the Dalitz plot conveys the important message that a
significant fraction
of $b\rightarrow c\bar cs$ processes could be seen in $D\overline D
\;\overline K X$.

The surplus of the inclusive kaon yield in $B$ decays
beyond all the conventional sources again indicates a significant
$B(\overline B\rightarrow D\overline D \;\overline K X)$ \cite{bdy2}.
The indication is further strengthened by the large observed
$K$-flavor
correlation with its parent $B$-flavor at time of decay
\cite{ktag,distinguish}.
The flavor of the kaon in $\overline B\rightarrow D\overline D
\;\overline K X$
is 100\% correlated with its parent $b$-flavor.
The momentum spectra of the inclusive $D$ yields in $B$ decays
indicates an excess of low momentum $D$'s over conventional
 sources \cite{lewis}.
A natural explanation can be found in $\overline
B\rightarrow D\overline D \;\overline KX$.

We are now ready to suggest several tests.
In addition to the ``indirect" measurement using
$B(b\rightarrow c\bar c s') \approx n_c -1 \;$
which involves large errors, we suggest to directly determine
$B(b\rightarrow
c\bar cs')$
by adding up the ``wrong-sign" charm in tagged $B$ decays
\cite{fwd,pub361},
\begin{eqnarray}
B(b\rightarrow c\bar cs') \approx B(b\rightarrow \bar c)  & = &
B(\overline B\rightarrow D^-_s X) +
B(\overline B\rightarrow \overline DX)+ B(\overline B\rightarrow
\overline{\Lambda}_c
X) + \nonumber \\
& + & B(\overline B\rightarrow \overline{\Xi_c}
X) + B(\overline B\rightarrow (c\bar c) X) \;.
\end{eqnarray}
The traditional lepton and $K^\pm$ tags could be supplemented by
other tags, such as $K^*$ and jet charge techniques.
Further, the number of $DD$ and $DD_s$ events per $\Upsilon
(4S)\rightarrow
B\overline B$ decay can be combined with the single, inclusive $D$
and $D_s$ yields
in untagged $B$ decay to determine $B(\overline B\rightarrow
\overline D X)$ and
$B(\overline B\rightarrow D^-_s X) $ \cite{fwd}.
Of course, $B^0 - \overline{B}^0$ mixing effects must be corrected
for
\cite{distinguish}. No tagging is required to measure $B(\overline
B\rightarrow (c\bar c)X)$.

\def\subdecay#1{\raise 5pt \hbox{\hspace {0.1 em} {\vrule height 15pt
depth -2.8 pt } \hspace {-0.75 em} $\longrightarrow #1$} }

A sizable $B(\overline B\rightarrow D\overline D \;\overline KX)$
would show up as a $D^{(*)}\overline K$ (from $cs$) enhancement.
The background at the $\Upsilon (4S)$ is much reduced because

\begin{eqnarray}
\Upsilon (4S) \rightarrow && \overline B B \rightarrow \overline D
\rightarrow
K \nonumber \\
&& \subdecay{D}\;,
\end{eqnarray}
which naturally yields a $DK$ correlation, while its $D\overline K$
correlation is suppressed.
The Dalitz plot allows to enhance the $D^{(*)}\overline K$ signal
correlation further by assuming
$$\frac{d\Gamma}{dm^2_{D^{(*)}\overline K}} \approx
\frac{d\Gamma}{dm^2_{cs}} \;.$$
The invariant mass spectrum of the $b\rightarrow c \bar c s$
transition
indicates that $D^{(*)}\overline K$ (from $cs$) tends to have a large
invariant mass, see Fig.~2.

The inclusive $D_s$ yield in B decays, $R_{D_s}$, has two roughly
equal
contributions. Figure 3 shows the measured momentum
spectrum
\cite{menary}.
Whereas the high peak is dominated by the exclusive two-body modes
$\overline B\rightarrow D^{(*)}D^{(*)-}_s$, the underlying dynamics
of the remainder
had been unclear.
The factorization assumption is successful in predicting ratios of
rates for the
two-body modes $\overline B\rightarrow D^{(*)} D^{(*)-}_s$
\cite{menary}.
Thus we assume factorization and predict that $b\rightarrow
c+D^{(*)-}_s$ is
dominated by the exclusive two-body decays $\overline B\rightarrow
D^{(*)}D^{(*)-}_s$ in
analogy to semileptonic decay of $B$ mesons. We calculate that
\begin{equation}\label{ddscds}
\frac{\Gamma (\overline B\rightarrow D^{(*)} D^{(*)-}_s )}{\Gamma
(b\rightarrow c+D^{(*)-}_s )}= 0.7 \pm 0.2 \;,
\end{equation}
where the quoted error refers to a variation in the $b$-quark mass,
$4.4 \leq m_b \leq 5.2 \; GeV$, and in the slope of the Isgur-Wise
function~\cite{cleo2}, $\rho^2 = 0.84 \pm 0.15$.
The numerator is the sum over the four exclusive two-body rates
obtained \cite{rosner,roberts} by using the heavy quark limit \cite{isgurwise}.
The denominator is the sum of two rates $b\rightarrow c+D_s$ and
$b\rightarrow c+D^*_s$.
It treats the $b\rightarrow c$ transition as if it were that of
free quarks \cite{palmerstech}. The decay constant $f_{D_s}$, the CKM elements
and the factorization parameter~\cite{bsw} $a_1$ all cancel in the ratio.  The
prediction Eq.~(\ref{ddscds}) can
currently be tested since the ratio
 $\Gamma (\overline B\rightarrow D^{(*)} D^{(*)-}_s ) / \Gamma
(b\rightarrow c+D^{(*)-}_s )$
 is an observable~\cite{nods} in which the uncertainty due to
$B(D_s \to \phi \pi)$ cancels.
The prediction Eq.~(\ref{ddscds}) together with the measured ratio
\cite{menary}
\begin{equation}
\frac{B(\overline B\rightarrow D^{(*)} D^{(*)-}_s )}{R_{D_s}} = 0.46
\pm 0.04\;,
\end{equation}
yields that~\cite{nods}
\begin{equation}
\label{dds}
B(\overline B\rightarrow DD^-_s X) \approx B(b\rightarrow
c+D^{(*)-}_s ) = (0.7 \pm 0.2) R_{D_s} \;.
\end{equation}
The remainder of the inclusive $D_s$ yield in $B$ decays
[$R_{D_s} - B(\overline B\rightarrow DD^-_s X) = (0.3 \pm 0.2) R_{D_s}$] could
be a significant fraction
of the
lower momentum $D_s$ mesons. One sizable source for it could be the
$b\rightarrow c\bar cs$
transition with $\bar ss$ fragmentation~\cite{bdy2},
\begin{equation}
B(b\rightarrow c\bar cs +\bar ss) \approx 0.01 - 0.03 \;.
\end{equation}
One generally expects one $D^-_s$ per such a transition, as long as
$D^{**-}_s$ and higher $D^-_s$ resonance production in
$b\rightarrow c\bar cs +\bar ss$ transitions is negligible.
The total $D^-_s$ production in flavor-tagged $\overline B$ decays is thus
expected to be
\begin{equation}
B(\overline B\rightarrow D^-_s X)\approx B(b\rightarrow c\bar cs
+\bar ss)
+ B(\overline B\rightarrow DD^-_s X) \approx 0.1 \;.
\end{equation}
The $D^+_s$ yield in flavor-tagged $\overline B$ decays is governed by the
$b\rightarrow c$ transition with $\bar ss$
fragmentation, and may be non-negligible
\begin{equation}
B(\overline B\rightarrow D^+_s X) =R_{D_s} -B(\overline B\rightarrow
D^-_s X) \sim 10^{-2} \;.
\end{equation}
For a model of the relative contributions to the $D^+_s$ yield from
$b\rightarrow c\bar ud, c\ell\nu , c\bar cs$ transitions with $\bar
ss$
fragmentation, we refer the reader to Ref. \cite{bdy2}.  The $D_s^+$
yield in flavor-tagged $\overline B$ decays has been
traditionally neglected~\cite{palmerstech,muheim,browder}.

In conclusion, by combining reliable theoretical calculations and
precise experimental measurements~\cite{pub361}, we
obtain a more accurate estimate of $B(b\rightarrow c\bar cs')$
and of $n_c$ than previous investigations
\cite{browder,muheim,baffle,fwd,bagan2,bagan3,voloshin}.
We predict
\begin{equation}
\label{bccsnc}
B(b\rightarrow c\bar cs') = 0.32 \pm 0.05  \;{\rm and}\;n_c = 1.30
\pm 0.05 ,
\end{equation}
which is significantly larger than the low experimental value
$\; n_c|_{exp} = 1.10 \pm 0.06$.  We believe that (\ref{bccsnc}) is on firm
ground, and expect an increase in the measured $n_c$ in the future.  Our
prediction can be tested in a variety of ways. First
we advocate
to measure $B(b\rightarrow c\bar cs')$ by counting up the number
of anticharmed hadrons (the ``wrong" charm flavor) per $\overline
B$-decay.
A sizable $BR$ for $\overline B \rightarrow D\overline D \;\overline
KX$
is our second prediction. It shows up as a large $\ell^-\;D$ and
$\ell^+\;\overline D$ correlation after removing $B^0 -
\overline{B}^0$ mixing effects~\cite{distinguish}, where the primary
lepton comes from one $B$ hadron and the charmed meson
from the other $B$-hadron in the event.  It can also be seen by observing
the
exclusive modes $\overline B\rightarrow D^{(*)}\overline{D}^{(*)}
\overline K$, and/or by searching for $D^{(*)}\overline K$ (from
$cs$)
correlations. There are many additional
implications,
consequences and tests which we hope to discuss in a forthcoming
report
\cite{bdy2}.

We are delighted to thank James D. Bjorken for insightful suggestions
and comments, and Martin Beneke, Tom Browder, Adrian Cho, Peter S.
Cooper, Amol Dighe, Estia Eichten, Scott Menary, Chris Quigg,
Jonathan L. Rosner and Alan Weinstein for discussions.
We thank Lois Deringer for typing this manuscript.  This work was
supported by the
Department of Energy, Contract. No. DE-AC02-76CH03000.

\begin{table}
\caption{The predicted semileptonic branching ratio, the
$B(b\rightarrow
c\bar cs')$ and $n_c$ taken from Bagan et al.~[6].}
\begin{tabular}{|c|c|c|c|}
Scheme & $B(\overline B\rightarrow X_c \ell\nu )$ & $ B(b\rightarrow
c\bar cs')$
& $n_c$  \\
\tableline
$\overline{MS}$ & 0.112 $\pm$ 0.017 & 0.35 $\pm$ 0.19 & 1.35 $\pm$
0.19 \\
\tableline
Pole mass & 0.120 $\pm$ 0.014 & 0.27 $\pm$ 0.07 & 1.27 $\pm$ 0.07
\end{tabular}
\end{table}

\begin{table}
\caption{The five classes of hadronization of $b\rightarrow c\bar
cs$.
$(c\bar c$) denotes charmonia not seen in $D\overline D X$, and class
(e)
involves $\bar ss$ fragmentation.}
\begin{tabular}{|c|c|c|c|}
Class & Mode & $BR$  & Reference \\
\tableline
(a) & $\overline B\rightarrow DD^-_s X$ & $ \; (0.7 \pm 0.2) R_{D_s}
\approx \;0.08$ &  \\
(b) & $\overline B\rightarrow \Xi_c \overline\Lambda_c X$ & 0.01 &
\cite{zoeller} \\
(c) & $\overline B\rightarrow (c\bar c) \overline KX$ & 0.03 &
\cite{browder,bdy2} \\
(d) & $\overline B\rightarrow D\overline D \;\overline KX$ & $\sim$
0.2 & \\
(e) & $b\rightarrow c\bar cs +\bar ss$ & $\sim few \times 10^{-2}$ &
\\
\tableline
Total: & $b\rightarrow c\bar cs$ & 0.31 $\pm$ 0.05 & \\
\end{tabular}
\end{table}

\begin{figure}
\centering
\mbox{\psfig{figure=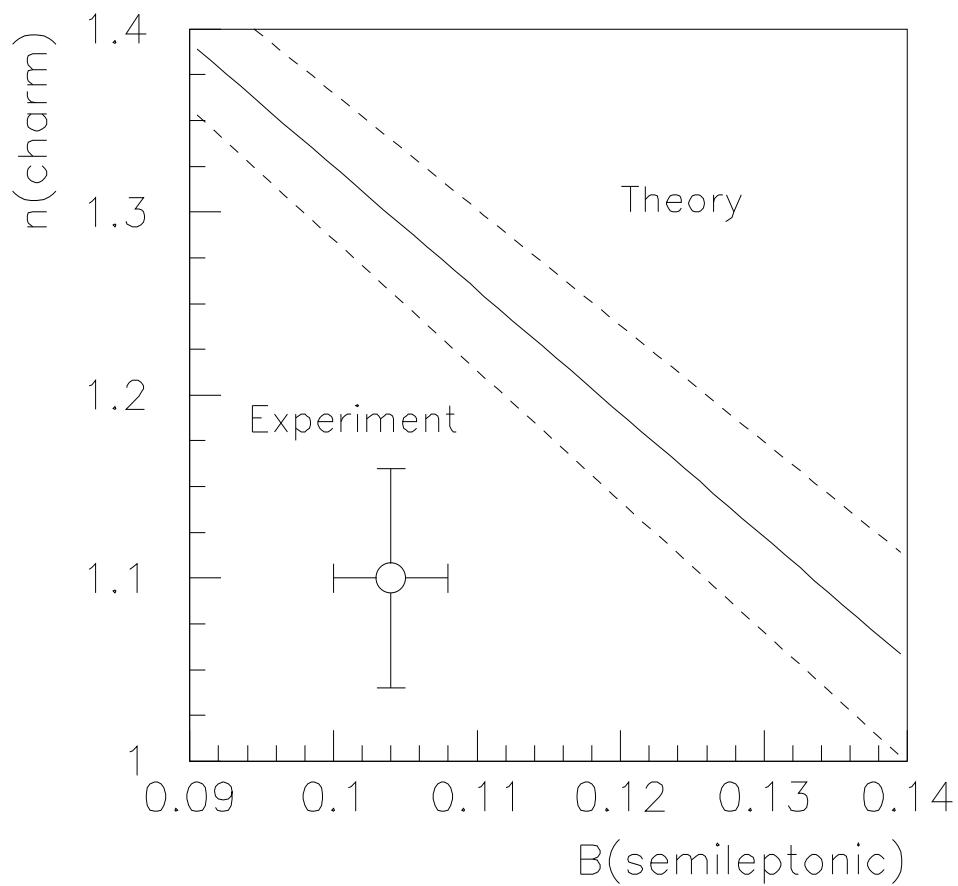,height=6.9in,width=5.6in}}
   \caption{Number of charm per $B$ decay ($n_c$) is plotted against the
$B$ meson
semileptonic branching ratio. The uncertainty in the theoretical prediction is
indicated by dashed lines.}
  \label{fg:ncbsl}
\end{figure}

\begin{figure}
\centering
\mbox{\psfig{figure=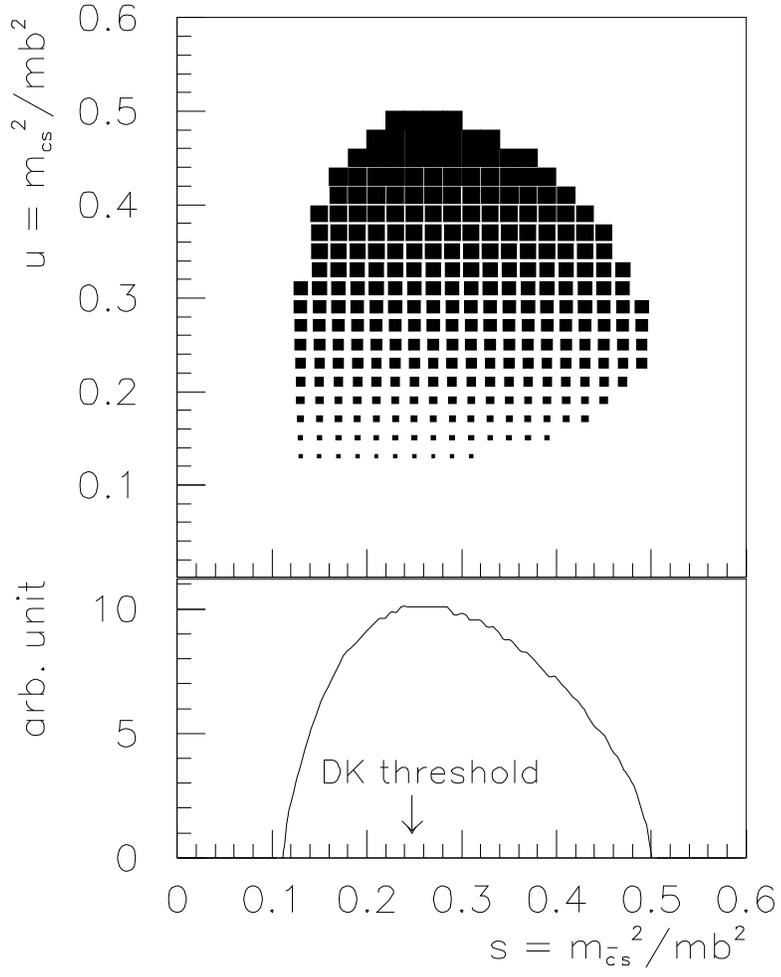,height=6.9in,width=5.6in}}
  \caption{Dalitz plot of the decay $b\to c \bar c s$ as a function of
    $u (= m_{c s}^2/m_b^2)$ and $s (= m_{\bar c s}^2/m_b^2)$. The
    projection onto
    the $s$ axis is shown at the bottom where
    the $\overline D \;\; \overline K$ threshold is indicated by an
    arrow.}
  \label{fg:dalitz}
\end{figure}

\begin{figure}
\centering
\mbox{\psfig{figure=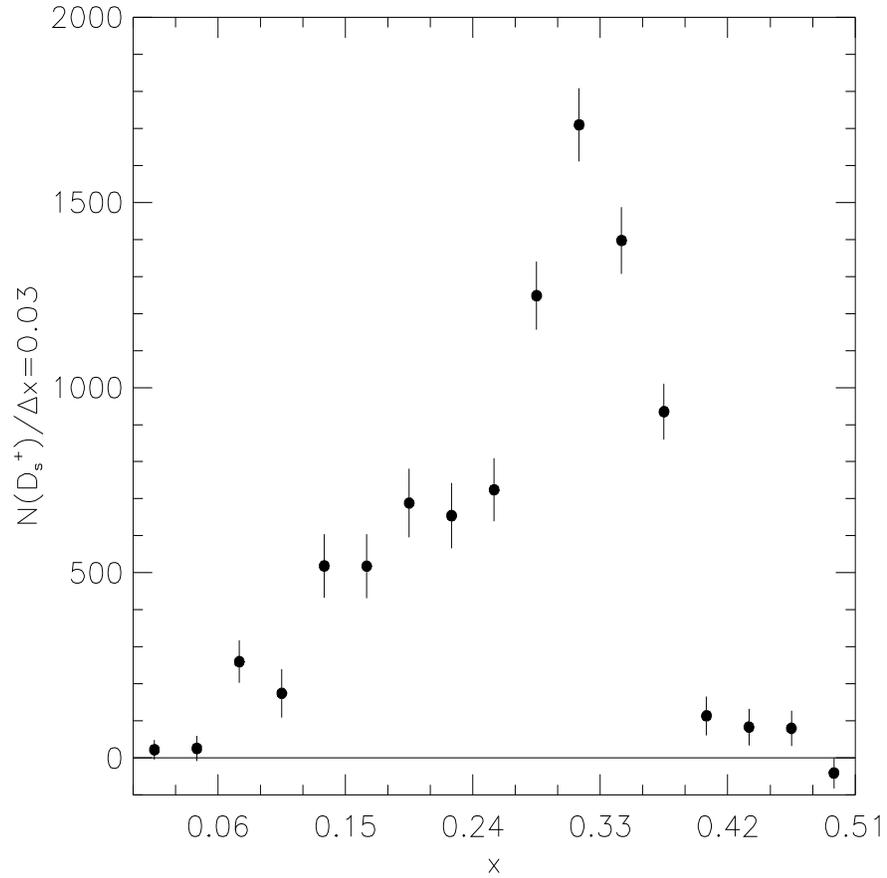,height=6.9in,width=5.6in}}
  \caption{Momentum spectrum of inclusive $D_s$ mesons produced in untagged
$B$ decays at the $\Upsilon (4S)$ as measured by the CLEO
   collaboration. The parameter $x$ is defined by
   $x = p_{D_s}/p_{\rm max}$ where
   $p_{\rm max}^2 \equiv E_{\rm beam}^2 - M_{D_s}^2$ .
   The continuum background has been subtracted.}
  \label{fg:dsmom}
\end{figure}


\begin{references}
\bibitem{br}
The inclusive yields of $D_s$ \cite{menary} and
$\Lambda_c$ \cite{zoellerthesis} in $B$ decay can be expressed in terms of the
$B(D_s\rightarrow \phi\pi)$
and $B(\Lambda_c \rightarrow pK^- \pi^+)$ as,
$$R_{D_s} = ( 0.12 \pm 0.01) \;\;\frac{0.035}{B(D_s\rightarrow \phi\pi)}
\;,$$
$$R_{\Lambda_c} = (0.041 \pm 0.008) \;\;\frac {0.044}{B(\Lambda_c
\rightarrow pK^- \pi^+ )} \;.$$
We choose the current central values $B(D_s \rightarrow \phi\pi )
=0.035$ and
$B(\Lambda_c \rightarrow pK^- \pi^+ )=0.044$. We alert the reader that
smaller
absolute $BR$'s for $D_s$ and $\Lambda_c$ decays increase
the yield of charm per $B$, and would
lessen the discrepancy between experiment and theory regarding $n_c$.
$B(\overline B\rightarrow \Xi_c \overline{\Lambda}_c X)$ is obtained
by combining $R_{\Lambda_c}$ and the relevant $\ell^\pm \Lambda_c$
measurement
\cite{zoeller} where the primary lepton comes from one $B$ and the
$\Lambda_c$ from the other $B$ in the $\Upsilon (4S)$ event.
The inclusive $BR$ into $(c\bar c)$ charmonia is obtained~\cite{bdy2}
to be $0.026 \pm 0.004$ by
summing over their observed and estimated $BR$'s which is larger than
previous estimates~\cite{browder}.
\bibitem{nokstar}A detailed analysis of inclusive $K^*$ yields in $B$
decays
suggests that the dominant source of $\overline K^*$ in
$\stackrel{(-)}{B}$ decays comes
from intermediate charmed hadrons. Thus there is not much room for
$\overline{K}^*$
in the process $\overline B\rightarrow D\overline D \;
\overline{K}^* X$.
\bibitem{nods}
Although $D_s^{**-}$ resonances are not normally seen as $D_s^- X$,
some are broad enough to be produced below the $\overline D^{(*)}
\overline K$ threshold where they decay to  $D_s^- X$.  That
contribution has not been included in the estimate for $\Gamma
(b\rightarrow c+D^{(*)-}_s )$.  We approximate
$\Gamma(b\rightarrow c+D^{(*)-}_s ) \approx
\Gamma(\overline B \to D D_s^- X)$, since $s\overline s$
fragmentation at the $ b\rightarrow c$ vertex is small.  Information
about this $s\overline s$ fragmentation comes from $\overline B \to
D_s^{(*)+} \overline K X \ell \nu$, which has not yet been observed
and for which upper limits exist~\cite{pdg}.  $B(\overline B \to
D_s^{(*)+} \overline K e \nu)$ is estimated to be $\sim \mbox{few} \times
10^{-3}$ [I.~Dunietz, J.L.~Goity, and W.~Roberts, in preparation].
\end{references}
\end{document}